\documentclass[twocolumn,showpacs,preprintnumbers,amsmath,amssymb]{revtex4}
\usepackage[dvips]{graphicx}
\usepackage{dcolumn}
\usepackage{bm}
\begin{document}
\title{A Josephson junction as a detector of Poissonian charge injection}
\author{J. P. Pekola}
\affiliation{Low Temperature Laboratory, Helsinki University of
Technology, P.O. Box 3500, 02015 HUT, Finland}


\begin{abstract}
We propose a scheme of measuring the non-Gaussian character of noise by a hysteretic Josephson junction in 
the macroscopic quantum tunnelling (MQT) regime. We model the detector as an (under)damped $LC$ resonator. It 
transforms Poissonian charge injection into current through the detector, which samples the injection 
statistics over a floating time window of length $\sim Q/\omega _{\rm J}$, where $Q$ is the quality factor of 
the resonator and $\omega _{\rm J}$ its resonance frequency. This scheme ought to reveal the Poisson 
character of charge injection in a detector with realistic parameters.
\end{abstract}
\maketitle

Presently there is considerable effort on characterizing and measuring the statistics of electrical current 
and the non-Gaussian nature of its fluctuations in mesoscopic conductors \cite{levitov96,nazarov03}. The 
discrete nature of charge injection, e.g., in tunnel junctions, with typically Poissonian statistics can be 
revealed by studying not only the average current $\langle I \rangle$ and its variance $\langle \delta I^2 
\rangle$, unlike in the case of normally distributed current, but higher moments also, most notably the third 
(central) moment $\langle \delta I^3 \rangle$. Normally these higher moments introduce very small signals and 
the filtering requirements are strict, because of which it is very hard to measure them (see, e.g., 
\cite{heikkila04,reulet03}). There is, however, one recent experiment which succeeded to demonstrate the 
existence of non-vanishing third moment in transport through a non-superconducting tunnel junction 
\cite{reulet03}. The measurement required very long averaging times. Therefore, alternative techniques to 
collect more information, and perhaps eventually to determine further higher moments, are definitely needed. 
In their recent article Tobiska and Nazarov \cite{tobiska03} proposed an overdamped Josephson tunnel junction 
(array) as a threshold detector to measure such full counting statistics (FCS) making use of rare 
over-the-barrier jumps arising from current fluctuations. The nearby on-chip detector is a definite benefit 
of this proposal due to its natural high bandwidth. Yet they considered the limit where tunnelling is 
perfectly suppressed whereby the set-up becomes experimentally less accessible. In our proposal we consider 
an underdamped single Josephson junction (or a DC-SQUID) in the MQT limit. We demonstrate that the 
experimental complications of the proposal \cite{tobiska03}, e.g., the need of a multijunction Josephson 
junction array, and the fact that the escape threshold is more difficult to measure in an overdamped 
junction, are overcome in our scheme, and show that the effect of higher moments is pronounced using a 
threshold detector with parameters deduced from earlier MQT experiments (see, e.g., Ref. \cite{balestro03}).

\begin{figure}
\begin{center}
\includegraphics[width=0.4
\textwidth]{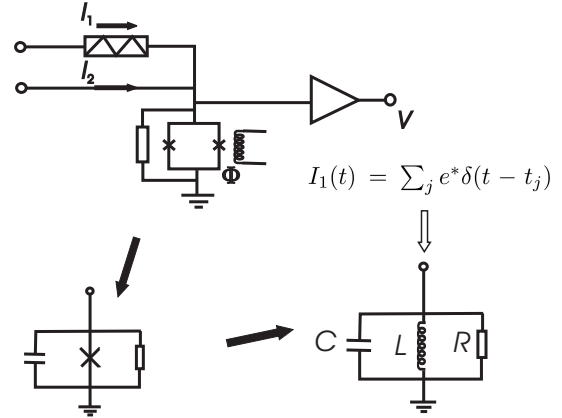}
\caption{The scheme of the threshold measurement and the model used to extract the results of this paper. The 
current through the Josephson junction, depicted as a DC-SQUID, is determined by the currents through the 
scatterer ($I_1$) and the ideal bias line ($I_2$), respectively. Switching from the supercurrent state to the 
normal state of the Josephson junction is signalled by a non-zero voltage $V$ at the output of the amplifier. 
The resonator model of the Josephson junction and the charge injection are discussed in the 
text.}\label{fig:scheme}
\end{center}
\end{figure}

We discuss a simplified model where charges are injected according to Poisson statistics on a Josephson 
junction, which in turn is described by a damped harmonic oscillator [$LCR$, or a linearized resistively and 
capacitively shunted junction ($RCSJ$) model]. We show how this environment performs a conversion from 
discrete (charge) statistics into continuous (current) statistics. The proposed measuring scheme is sketched 
in Fig. \ref{fig:scheme}. There are two injecting lines, with currents $I_1$ and $I_2$. $I_1$ is generated by 
a voltage bias across the scatterer (for example a tunnel junction), and $I_2$ runs in a directly connected 
line to be discussed below. We neglect the influence of the transmission line connecting the injecting lines 
and the measuring Josephson junction. Capacitance of the injecting junction can be included in the $RCSJ$ 
capacitance, and similarly the additional injection line ($I_2$) and the connection to the voltage amplifier 
can be modelled as a parallel inductance, which may reduce the resolution of the detector as will be 
discussed at the end. For practical implementations we depict the detector as a DC-SQUID with critical 
current tunable by magnetic flux $\Phi$.

The detector is driven by elementary charges $e^*$ "dropped" on the $LCR$ resonator according to Poisson 
statistics. Therefore, each event poses an elementary current $e^*\delta(t-t_j)$ into the resonator such that  
$I_1(t)=\sum _j e^*\delta(t-t_j)$ as illustrated in Fig. \ref{fig:scheme}. Here $t_j$ is the time instant of 
the $j$th event and $\delta(t)$ is the Dirac delta function. (Alternatively we could consider a voltage step 
of height $e^*/C$ at each instant $t_j$.) Our task is to evaluate the current $I$ and its moments through the 
junction (inductor $L$). The elementary current $i(t,t_j)$ through the Josephson junction at the time instant 
$t$, due to this $j$th event can be easily solved, and is given in the case of an underdamped resonator 
(quality factor $Q\equiv R/\sqrt{L/C} > 1/2$) by
\begin{equation} \label{eq1}
\begin{split}
i(t,t_j)= & e^* \omega _{\rm J}(1-\frac{1}{4Q^2})^{-1/2}\exp [-\frac{1}{2Q}\omega _{\rm J}(t-t_j)] \\ & \sin 
[(1-\frac{1}{4Q^2})^{1/2}\omega _{\rm J}(t-t_j)]\theta (t-t_j).
\end{split}
\end{equation}
Here, $\omega _{\rm J} = (LC)^{-1/2}$ is the plasma frequency of the junction and $\theta(t)$ is the 
Heaviside step function. Analogous results for overdamped case ($Q < 1/2$) exist, but they will not be 
considered explicitly here, since the detector is assumed to {\sl switch} from supercurrent state to 
resistive state. Therefore we assume $Q > 1/2$ in what follows unless otherwise noted. It is interesting to 
see some properties of these elementary oscillations. At the time instant $t_j$ all charge $e^*$ is stored in 
the capacitor $C$, whereafter it performs damped oscillations at (angular) frequency $\omega _{\rm J}$. On 
integrating $i_j(t)$ one finds a consistent result: $\int _{-\infty}^{+\infty}i(t,t_j)= e^*$.

\begin{figure}
\begin{center}
\includegraphics[width=0.5
\textwidth]{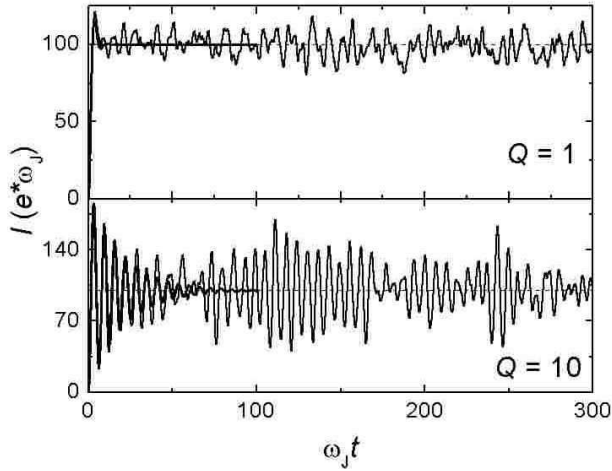}
\caption{Simulated $I(t)$ (thin solid lines) with the following parameters. $\bar{I} = 100$, $p=0.01$ and 
with two values of quality factor, $Q=1$ and $Q=10$, respectively. Thick solid lines are expectations of 
$\langle I(t)\rangle$ according to Eq. (\ref{eq4}).}\label{fig:sim}
\end{center}
\end{figure}
$I_1(t)$ induces current $I(t)$ through the Josephson junction that can be expressed as the sum of the 
elementary currents of Eq. (\ref{eq1}) on the basis of the linearity of the equation whose solution it is:
\begin{equation} \label{eq2}
I(t)= \sum _j i(t,t_j),
\end{equation}
where now the time instants $t_j$ are perfectly uncorrelated. One can numerically simulate the current 
following Poisson principle: divide time into very small intervals $\delta t$ and set the probability $p \ll 
1$ for one charge to tunnel during this interval. No multiple events in any of the $\delta t$ intervals are 
allowed. The average current $\bar{I}$, $p$ and $\delta t$ are related through $\bar{I} = pe^*/\delta t$. 
Figure \ref{fig:sim} illustrates results of simulation with two different values of $Q$, time unit is $\omega 
_{\rm J}^{-1}$, and $\bar{I} = 100$, in units $e^* \omega _{\rm J}$. The initial rise and oscillations are 
due to the fact that the injection of charges is suddenly initiated at $t=0$. At larger values of $Q$ the 
junction tends to oscillate at (angular) frequency $\omega _{\rm J}$, but it is driven by random events and 
thereby dephased.

Simulation yields a representative example based on the random numbers determining the time instants $t_j$. 
It is more general and interesting to investigate the properties of the distribution of current. Let us first 
consider the various moments of current $I(t)$. This can be done in the following way. We take a long enough 
time interval $\tau$ such that the interesting instant $t$ at which we want to evaluate the moments of 
current satisfies $t \le \tau$. Now we consider different ensembles of instants $t_j$, such that $N$ charges 
are injected within $\tau$, and then weight all these configurations by the Poisson probability $P_{\rm 
Poisson}(N)$. Since all the $j=1,...,N$ events are uncorrelated and evenly distributed over $0 \le t_j \le 
\tau$, we may write for the $n$th moment of $I(t)$:
\begin{equation} \label{eq3}
\begin{split}
\langle I^n(t) \rangle = & \sum _{N=1} ^{\infty} P_{\rm Poisson}(N) \\ & \tau ^{-N}\int _0 ^\tau \int _0 
^\tau \cdot \cdot \cdot \int _0 ^\tau dt_1dt_2 \cdot \cdot \cdot dt_N [\sum _{j=1} ^N i(t,t_j)]^n.
\end{split}
\end{equation}
It is straightforward to integrate for $\langle I^n(t) \rangle$ using $i(t,t_j)$ of Eq. (\ref{eq1}). Below we 
summarise results for the three lowest moments of $I(t)$. The average current reads
\begin{equation} \label{eq4}
\begin{split}
& \langle I(t) \rangle = \bar{I}\lbrace 1-\exp(-\frac{\omega _{\rm J}}{2Q}t)\\ & 
[\cos(\sqrt{1-\frac{1}{4Q^2}}\omega _{\rm 
J}t)+\frac{1}{\sqrt{1-\frac{1}{4Q^2}}}\sin(\sqrt{1-\frac{1}{4Q^2}}\omega _{\rm J}t)]\rbrace.
\end{split}
\end{equation}
Here we have identified $\bar{I} = e^*\langle N \rangle/\tau \equiv \frac{e^*}{\tau}\sum _{N=1}^{\infty} N 
P_{\rm Poisson}(N)$, which is the asymptotic value of $\langle I(t) \rangle$ on $t \rightarrow \infty$. The 
thick lines in Fig. \ref{fig:sim} show the result of $\langle I(t) \rangle$ using Eq. (\ref{eq4}), which 
indeed seem to follow the mean of the simulated curves as expected. In what follows, we drop out the argument 
$t$, and consider only results after the initial transient, i.e., $t \gg 2Q/\omega_{\rm J}$. The
second raw moment reads
$\langle I^2 \rangle = \frac{Qe^*\omega _{\rm J}}{2}\bar{I} + \frac{{e^*}^2}{\tau^2}\langle N(N-1)\rangle$,
where $\langle N(N-1)\rangle \equiv \sum _{N=1} ^{\infty} N(N-1)P_{\rm Poisson}(N)$. The more interesting 
second central moment, the variance $\langle \delta I^2 \rangle \equiv \langle (I-\langle 
I\rangle)^2\rangle$, then reduces to
\begin{equation} \label{eq6}
\langle \delta I^2 \rangle = \frac{Qe^*\omega _{\rm J}}{2}\bar{I}.
\end{equation}
After a straightforward derivation we similarly obtain the third central moment, $\langle \delta I^3 \rangle 
\equiv \langle (I-\langle I\rangle)^3\rangle$, reading
\begin{equation} \label{eq7}
\langle \delta I^3 \rangle = \frac{2}{3(1+2/Q^2)}(e^*\omega _{\rm J})^2\bar{I}.
\end{equation}
According to Eq. (\ref{eq6}), the shot noise of the injecting junction, $2e^*\bar{I}$, is amplified by the 
quality factor of the resonator over the band whose width is $\sim \omega _{\rm J}$, the (maximum) response 
frequency of the detector. A measure of the non-Gaussian character of the current distribution is its 
skewness \cite{mathw}, defined as $S=\langle \delta I^3 \rangle/\langle \delta I^2 \rangle^{3/2}$, which, 
according to Eqs. (\ref{eq6}) and (\ref{eq7}) reads 
\begin{equation} \label{eq8}
S=\frac{2^{5/2}}{3(Q^{3/2}+2Q^{-1/2})}(e^*\omega _{\rm J})^{1/2}\bar{I}^{-1/2}.
\end{equation}
Results (\ref{eq6})-(\ref{eq8}) hold also in the overdamped case ($Q < 0.5$). It is interesting to note some 
general features of $S$ in Eq. (\ref{eq8}). The non-Gaussian "strength" increases, in accordance with the 
central limit theorem, with decreasing $\bar{I}$ (less events recorded). The detector exhibits a memory of 
events over a time $\sim 2Q/\omega _{\rm J}$ in the underdamped [$1/(Q\omega _{\rm J})$ in the overdamped] 
case, and therefore, the skewness attains its maximum value close to the crossover between underdamped and 
overdamped behaviour, $Q \simeq 1$: here the memory of the detector is shortest, and it responds to only a 
small number $\bar{n}$ of Poisson distributed events through the scatterer. In the example discussed below 
$\bar{n}\sim (\bar{I}/e^*)(2Q/\omega _{\rm J}) \simeq 40$.

\begin{figure}
\begin{center}
\includegraphics[width=0.5
\textwidth]{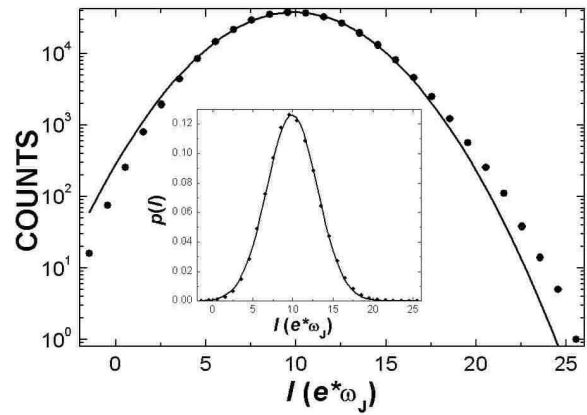}
\caption{Simulated current distribution at $\omega _{\rm J}t = 40$, $Q=2$ and $\bar{I}=10e^*\omega _{\rm J}$. 
The main frame shows the number of counts (out of 300 000) in solid dots, and the solid line is the best 
Gaussian fit to it. The inset shows the same results but as the true distribution, = counts/300 000, and on 
linear scale.}\label{fig:distr}
\end{center}
\end{figure}  
Figure \ref{fig:distr} shows an example of the simulated ($3\cdot 10^5$ repetitions) current distribution at 
the time instant $\omega _{\rm J}t = 40$ (far enough after the initial transient, $\omega _{\rm J}t \gg 2Q$) 
for the injected current with $\bar{I}=10e^*\omega _{\rm J}$ and for a detector whose $Q=2$. The main frame 
with logarithmic vertical scale shows the number of counts (in solid circles), demonstrating how the 
simulation differs from the Gaussian fit shown by the solid line. There are more hits at high currents in the 
simulation as compared to the normally distributed events as one would expect. The same is shown as the true 
distribution (counts divided by $3\cdot 10^5$) on linear scale in the inset. Table \ref{tab:1} gives a 
comparison between the simulated values and the theoretical predictions [Eqs. (\ref{eq4})-(\ref{eq7})] of the 
three lowest moments. The correspondence is satisfactory, although the variance $\langle \delta I^2 \rangle$ 
falls outside the $1\sigma$ uncertainty margin. 
\begin{table}
\caption{Results}
\label{tab:1}
\begin{tabular}{lll}
\hline\noalign{\smallskip}
 & simulation & theory \\
\noalign{\smallskip}\hline\noalign{\smallskip}
$\langle I\rangle$ & 10.01$\pm$ 0.01 & 10.00\\
$\langle \delta I^2 \rangle$ & 9.91$\pm$ 0.03 & 10.0\\
$\langle \delta I^3 \rangle$ & 4.65$\pm$ 0.22 & 4.44\\
\noalign{\smallskip}\hline
\end{tabular}
\end{table}

\begin{figure}
\begin{center}
\includegraphics[width=0.4
\textwidth]{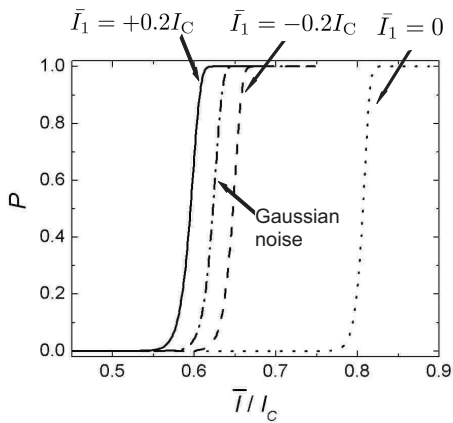}
\caption{Escape probability $P$ of a Josephson junction as a function of the average current $\bar{I}$ 
through the detector under different noise conditions. The shift between the solid and the dashed lines 
arises from Poisson statistics of charge injection. The dash-dotted line is the escape probability with the 
corresponding Gaussian noise (same $\langle \delta I^2\rangle$). Escape in the ideal case of noiseless 
current is shown by the dotted line.}\label{fig:hist}
\end{center}
\end{figure}

Next we make a judgment of whether such a threshold detector provides a viable means to measure FCS. To this 
end we need to consider the escape rates from the supercurrent state. We assume low temperature $T$ such that 
thermally activated switching is suppressed. This is the case when $T < \hbar \omega _{\rm J}/(2\pi k_{\rm 
B})$, which is an easily accessible regime experimentally \cite{martinis87}. Escape rate in the MQT regime in 
the presence of Gaussian noise has been discussed, e.g., in Ref. \cite{martinis88}. Here, we will allow for 
more general current statistics. Using the standard decay law we find that the probability of escape to the 
resistive state is given by 
\begin{equation} \label{eq9}
P = 1-\exp[-\int _{t_0}^{t_0+\Delta t}\Gamma(I(t))dt],
\end{equation}
where $\Delta t$ is the duration of the current pulse starting at $t=t_0$ over which we monitor escape 
statistics. $\Gamma$ is the current dependent escape rate in the MQT process for which one can find explicit 
expressions that depend on the junction and circuit parameters \cite{caldeira81}: $\Gamma = A \exp(-B)$, 
where $A=\chi \sqrt{\hbar \omega _{\rm J}\Delta U} /(2\pi\hbar)$ and $B=s \Delta U/(\hbar \omega _{\rm J})$. 
$\Delta U$ is the $I$ dependent barrier height, and parameters $\chi$ and $s$ are $Q$-dependent. They assume 
values $\chi=12 \sqrt{6\pi}$ and $s=36/5$ for large $Q$. For a pulse with $\Delta t \gg Q/\omega _{\rm J}$ we 
may write $\int _{t_0}^{t_0+\Delta t}\Gamma(I(t))dt\simeq \langle \Gamma \rangle \Delta t$, where $\langle 
\Gamma \rangle \equiv \int _{-\infty}^{+\infty} \Gamma(I)p(I)dI$. Here $p(I)$ is the current distribution 
approximated, e.g., in Fig. \ref{fig:distr}. In the scheme of Fig. \ref{fig:scheme} the average current 
$\bar{I}$ through the Josephson junction can be generated by any combination of the two (average) currents 
$\bar{I_1}$ and $\bar{I_2}$ with the constraint $\bar{I}=\bar{I_1} + \bar{I_2}$. Of particular interest are 
the cases where $\bar{I_1}$ is either $=0$, or it has positive or negative values of equal magnitude. 
Difference in the escape characteristics between the latter two cases provides a measure of the asymmetry of 
$p(I)$ around its mean, the central topic of this article. Figure \ref{fig:hist} shows the escape histograms 
calculated under such three conditions using trapezoidal current pulses of duration $\Delta t = 100$ $\mu$s 
\cite{balestro03,pulse}. We assume that the detector junction has a critical current $I_{\rm C}=1$ $\mu$A 
[$L=\hbar/(2eI_{\rm C})$], $C=0.1$ pF, and other parameters and $p(I)$ are as in Fig. \ref{fig:distr}. (With 
these parameters $\Delta t \gg Q/\omega _{\rm J}$.) We assume pure MQT escape with low dissipation 
\cite{caldeira81,martinis88}. The histograms are plotted as a function of average current $\bar{I}$ through 
the detector driven by the two injection currents in different proportions. The histogram shown in dotted 
line corresponds to no current fluctuations, i.e., all current is driven through the ideal line ($\bar{I_1} 
=0$). The solid line is for the case when current through the scatterer and that through the detector point 
in the same direction ($|\bar{I}|=|\bar{I_1}|+|\bar{I_2}|$). With our circuit parameters $\bar{I_1}=+0.2 
I_{\rm C}$. The dashed line is for the case when current $I_1$ points opposite to that through the detector 
($|\bar{I}|=\big{|}|\bar{I_1}|-|\bar{I_2}|\big{|}$ and $\bar{I_1}=-0.2 I_{\rm C}$). The average shift of the 
$\bar{I_1} = \pm 0.2 I_{\rm C}$ histograms with respect to the dotted line is due to the variance of current, 
whereas the pronounced shift between the last two is the more interesting effect of non-Gaussian current 
statistics.

We conclude with a few practical remarks. The scheme presented here is simplified in many ways. Firstly, we 
do not take into account the (weak) dependence of the plasma frequency (Josephson inductance) on $I$ to keep 
the discussion more phenomenological and transparent. This is fairly well justified for example in case of 
Fig. \ref{fig:hist}, because all the relevant escape currents are smaller than $0.8I_{\rm C}$, and $\omega 
_{\rm J}\propto (1-I/I_{\rm C})^{1/4}$. Secondly, the escape histograms calculated assume low dissipation, 
which is not truly the case. The influence of dissipation on the MQT rate through environmental noise can be 
taken into account by a minor scaling of $\chi$ and $s$ parameters \cite{martinis88}, and again to keep 
analysis on the basic level, we omit this since the effect is weak even when $Q=2$. Thirdly, we assume that 
the injected charges do not produce current pulses either in the $I_2$ current line, nor in the line to the 
voltage amplifier. This can be realised by large inductance ($L_{\rm ext}$) in these lines, which in practice 
means long and narrow wires. If $L_{\rm ext} \gg L$, our argument is justified. Finally, the presented model 
is based on classical description of the circuit dynamics: this is a valid starting point in the case of a 
tunnel junction scatterer whose tunnel resistance $R_{\rm T} > \hbar/e^2$. 

Very valuable discussions with Dmitri Averin, Frank Hekking, Rosario Fazio, Tero Heikkil\"a and Antti 
Niskanen are gratefully acknowledged.


\begin{thebibliography}{99}

\bibitem{levitov96} L. S. Levitov, H. W. Lee, and G. B. Lesovik, J. Math. Phys. {\bf 37}, 4845 (1996).

\bibitem{nazarov03} {\sl Quantum Noise in Mesoscopic Physics}, edited by Yu. V. Nazarov (Kluwer, Dordrecht, 
2003).

\bibitem{heikkila04} T. T. Heikkil\"a and L. Roschier, submitted (2004).

\bibitem{reulet03} B. Reulet, J. Senzier, and D. E. Prober, Phys. Rev. Lett. {\bf 91}, 196601 (2003); B. 
Reulet, L. Spietz, C. M. Wilson, J. Senzier, and D. E. Prober, cond-mat/0403437.

\bibitem{tobiska03} J. Tobiska and Yu. V. Nazarov, cond-mat/0308310 (2003). 

\bibitem{balestro03} F. Balestro, J. Claudon, J. P. Pekola, and O. Buisson, Phys. Rev. Lett. {\bf 91},158301 
(2003).

\bibitem{mathw} See, e.g., Eric Weisstein's World of Mathematics, http://mathworld.wolfram.com

\bibitem{martinis87} See, e.g., J. M. Martinis, M. H. Devoret, and J. Clarke, Phys. Rev. B {\bf 35}, 4682 
(1987), or Ref. \cite{balestro03}.

\bibitem{martinis88} J. M. Martinis and H. Grabert, Phys. Rev. B {\bf 38}, 2371 (1988).

\bibitem{caldeira81} A. O. Caldeira and A. J. Leggett, Phys. Rev. Lett. {\bf 46}, 211 (1981); Ann. Phys. 
(N.Y.) {\bf 149}, 374 (1983); {\bf 153}, 445 (1984).

\bibitem{pulse} We thus assume that $\bar{I}=$ constant for $t_0 < t < t_0 + \Delta t$, and that it vanishes 
otherwise. The transients at the beginning and at the end of the pulse are, however, assumed to be adiabatic, 
i.e., their durations are $\gg \omega _{\rm J}^{-1}$. 



\end{thebibliography}
\end{document}